# A FRAMEWORK FOR 3D INTERACTION TECHNIQUES

Pablo Figueroa
pfiguero@cs.ualberta.ca
Ph.D. Student
Computing Science Department
University of Alberta
Edmonton, Alberta. Canada
T6H 4N1

Mark Green
smmark@cityu.edu.hk
Professor
School of Creative Media
City University of Hong Kong
China

Benjamin Watson
watsonb@cs.nwu.edu
Assistant Professor
Department of Computer Science
Northwestern University
Evanston, IL. United States of America
60201

## ABSTRACT

This paper presents a software architecture for 3D interaction techniques (ITs) and an object oriented, toolkit-independent framework that implements such architecture. ITs are composed of basic *filters* connected in a dataflow, where *virtual input devices and objects in the scene* are sources of information. An *execution model* defines the general flow of information between filters. This framework has been designed to be extensible: new information types, new input devices, new execution models, or new interaction techniques can easily be added. Application specific code and application specific ITs are seamlessly integrated into this architecture.

Applications that use this framework are designed and written in two standard languages: UML and XML. XML documents are parsed to generate code for a particular VR toolkit or environment.

## INTRODUCTION

Moller and Haines [20, pp.8-12] describe the architecture of a real-time graphics application as a pipeline with three main stages: application, geometry, and rasterizer. The architecture for the last two stages is widely known, but little research has been done in the first one, where interaction techniques take place. This paper presents an architecture and a object oriented framework, ITLib[1], that organize and compose behavior in the application stage.

The architecture is based on the Pipes and Filters architecture [8], in which sources of information generate a flow of information that is propagated to interconnected filters. Its design fulfills the following requirements:

A. ITs are reusable. Each technique is an identifiable and separate object that can be used in new applications and can be collected in a library.
B. ITs are interchangeable. It is possible to exchange one technique for another, for prototyping and testing purposes.
C. New devices, and the data types of the information they generate, can be added to the environment. ITs that receive predefined data types can be connected to new devices, so they are independent of physical devices.
D. Application specific code can be added to the environment as new filters. This code can become new ITs.
E. ITs are independent of any particular graphics API. They use generic wrappers [12, p. 139] to communicate with the objects in the scene.

The following sections present previous work, the basic concepts in the framework, some use case scenarios and examples, the current implementation, and some conclusions.

## RELATED WORK

Several mechanisms have been proposed in the VR community to describe and implement interaction techniques. Bowman [4] classifies ITs in families and defines a framework to construct new ones, but without a particular mapping to actual code in an application. Steed [27] provides an environment for designing 3D interaction techniques by manipulating graphical objects; however, the design of an entire application can be difficult to understand because of the visual complexity, and there are no mechanisms for encapsulation and abstraction of ITs. Halliday [14] defines a language to describe objects and animations, and a menu based interface to create animations by connecting input device events with some behaviors. However, the possible behaviors are rather simple, and there are few possibilities for composing. Broll [7] proposes a scripting language to describe all elements in a VR environment, including ITs, with an implicit execution model behind it. This work is similar to his, with the advantages that we define extension mechanisms as part of ITLib. The most complete work in this area is by Jacob [19], who presents a low-level specification language for ITs, an implementation of an editor for ITs, and a runtime environment. Our model has the same expressive power that his language has, but manipulates elements at a higher level of abstraction, has mechanisms that relate physical devices to ITs, enables encapsulation and abstraction of ITs, allows changing ITs, and provides an extensible execution model. However, our model doesn't have a distributed implementation.

---

[1] ITLib also refers hereafter to the library of interaction techniques that implements the framework.

Current VR toolkits offer limited support for describing ITs and adding new input devices. They usually provide callbacks or events from a fixed set of devices and a predefined execution model. Some VR toolkits, such as [22], rely on other APIs to get information from devices. Others, such as [1, pp. 20-22] [2] [16] [25] [10], define a fixed set of events that can be captured by user-defined callbacks, although it is not clear how to add new VR devices or new events. Some offer a physical model of devices [29]. Only some, such as [2, pp. 17-19] [3] [30] [13] [18], offer an abstract and extensible model for devices like the one used in ITLib.

Most VR toolkits don't model ITs as primary objects, so code for each IT is fragmented inside different callbacks, making them difficult to reuse. MRObjects [13] defines a hierarchy of classes for ITs that can be used to create more complicated ones, but the order of execution of ITs is unspecified. SVIFT [15] provides a library of interaction techniques as C components, but it is based in the limiting event model, reuse capabilities in C are limited, and it is not clear how complex ITs can be composed. Java3D [18] defines an abstract class called *Behavior* that can be used to create concrete ITs, and it is possible to define an order of execution between them, although there are no ITs implemented in this API.

Standard WIMP-based interfaces are based mainly in the event model of execution, where callbacks respond to a set of predefined events. Some authors [21] believe this model is not sufficient for VR applications, and new models have to be proposed to deal with problems of multimodality and synchronization. Some models have been proposed to deal with this issue [28] with limited success.

In summary, this work introduces the concept of a library of interaction techniques based on ideas from the systems mentioned above. We add in this library encapsulation, extension, and abstraction mechanisms for ITs and devices, clear interfaces between physical devices and ITs, and a novel way to separate the execution model from the application code that allows us to plug in different alternatives.

## BASIC CONCEPTS

At a low level, ITLib is composed of *filters* that can be connected in different ways in a dataflow arrangement. Input devices are modeled as a set of *virtual input devices* (a subclass of filter), which generate different types of information. An *IT* is modeled as a set of filters connected in a small dataflow. An *Execution Model* defines how the overall dataflow is executed. Finally, there are wrappers for the *VR Toolkit* in use and for the *objects in the scene*, allowing us to connect ITs to the rest of the VR environment in an standard, toolkit-independent way. These concepts are described in more detail in the following paragraphs and examples are given in a following section.

Ideally, an application using this framework runs at least two threads (Figure 1): One for reading and queuing information from physical input devices (Thread 1) and one for the application's main loop that executes the dataflow of ITs and renders the scene (Thread 2). Thread 1 has routines to read input devices and either queue events or keep the last one. The execution model in Thread 2 reads the queued information from Thread 1 and propagates all messages throughout the filters. It is the responsibility of the application's main loop to execute one step of the execution environment between renderings.

A *filter* (Figure 2) is the smallest process unit in the dataflow, composed of input and output ports, shared information (i.e. objects in the scene), and state information about input in the past. An *input port* is a part of a filter used to receive and collect information of a particular type. An *output port* collects the information generated and sends it to all filters that follow, called its listeners[2].

---

[2] Listeners are all filters interested in an object's changes. It can be understood as the Observer pattern in [12].

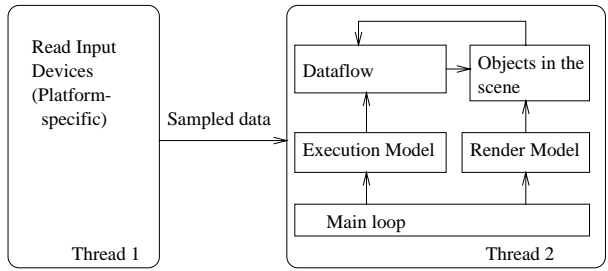

**Fig. 1** Execution view of the model.

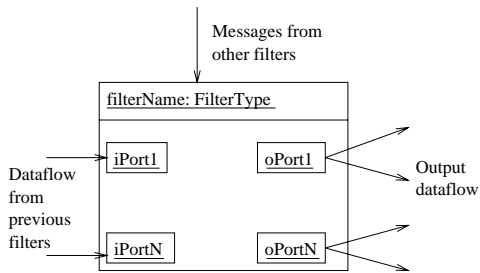

**Fig. 2** General Presentation of a filter in UML.

The filter's internal behavior is defined by the following three methods, which can be separately called by the Execution Environment:
A. Collect Input Information. The filter processes information from its input, deleting redundant information, for example.
B. Process Input. This method computes the information sent by the output ports and sends messages to other filters, when necessary, in order to change their behavior.
C. Send Output. The output generated by the filter is sent to its successors.

State changes in objects in the scene are not executed until the end of a step, ensuring consistency. A filter can translate input from a new device to an old filter, control the behavior of several filters by sending messages to them [3], or compute a new value from its input.

The basic types of information allowed in the dataflow are taken from [11], but they can be extended as necessary: Locator (a 6DOF value), Valuator (a real value), Pick (a selection from a set of options), and Button. We simplify Jacob's model of continuous and discrete variables to a discrete gathering of sampled information from devices or objects in the scene.

---

[3] Control filters can model the constraint's behavior in Jacob's [19, p. 7] model.

A *virtual input device* is a particular type of filter with no input. It uses a platform-specific implementation to get sampled data from a physical device, and it translates this information to allowed data types in the dataflow.

The *execution environment* defines the order of execution of filters in the dataflow. In the most simple one, which is used in VRML environments [9], an input is propagated throughout all filters in the dataflow before the next input is considered. We are working on new execution environments that parallelize the dataflow and place some run-time constraints on the filters, in order to achieve better performance and a fixed output frame rate.

An *interaction technique* is a collection of interconnected filters that performs a particular task. Its methods are the same as the filter's, therefore it can encapsulate the complexity of its implementation without losing uniformity. Its input and output ports are a subset of those in its filters.

*Wrappers for objects in the scene* [4] allow a filter to communicate with objects in a pre-defined and toolkit-independent manner. Their methods can be classified as:
A. Object Manipulation Methods. an IT requires standard ways to obtain information from and change the state of each object.
B. Listener Registration. An object can generate information each time its state changes. These methods allow different filters in the dataflow to get feedback from the objects in the scene.

The *wrapper for the VR toolkit* [5] comprises the set of standard services expected from the virtual reality toolkit. These services include access to the scene graph, the current viewpoint, the objects in the current frustum, and all available input devices.

## USE CASE SCENARIOS AND EXAMPLES

### (1) IT Definition and Encapsulation

We now show how a 3D IT can be expressed in terms of the previous basic concepts, how ITs can be connected to develop an application, and how an IT hides the complexity of its implementation.

---

[4] For simplicity's sake, we omit the word wrapper in the following sections.

[5] Hereafter VR Toolkit.

In order to define a new IT, the following steps have to be performed:
A. Describe the input of the IT in terms of data required from input devices and changes in the scene that the technique needs to know about.
B. Describe the expected output from the IT in terms of changes in the scene throughout the interaction. Define states in the interaction, if necessary.
C. Look for filters or ITs in ITLib that can be adapted for particular needs, by changing connections or behavior.
D. Define any additional filters required. Create control filters to switch between the states in the interaction.
E. Interconnect virtual input devices, objects in the scene, and filters.
F. Encapsulate filters in an IT object. Define the IT input and output as a subset of the ones in its filters.

We'll use these basic filters in the following examples:
MoveByLocator: Its input ports receive the object obj to move and a Location pos. It moves obj according to the changes of pos. Changes in pos can be taken as either a new absolute position of an object, or a new offset from the object's current position. It doesn't have output ports.
SelectObj: Superclass of any selection. It has an output port that sends selected objects and no input ports[6].
Select1ByPointing: Its input port is a location pos, defining a position in 3D space and an orientation. Its state is the current set of select-able objects objs. It computes an object pointed by the locator pos and sends it through its output port.
Select1ByTouching: Its input port is a location pos. Its state is the current set of select-able objects objs and the current object in the scene representing the user's hand. It calculates an object colliding with the hand representation and sends it through its output port.
ChangeObject: Its input port receives objects. Its state is the last object received. It changes a characteristic of the object, for example if the bounding box is visible. It doesn't have output ports.

Our first example is the Go-Go Technique [23], an IT to lengthen the user's virtual arm for reaching distant objects. GoGoIT is defined as follows:

A. Its input is the position of the user's hand and head (both of type Locator)

---

[6] This output can be seen as a Pick, according to Foley's taxonomy [11].

B. Its output is a selected object from the scene. The interface has to change to show the current state of the interaction technique, as follows:
  1) A virtual hand is moved according to the user's hand position and the GoGo lengthening scheme.
  2) When lengthening, a cube shows the position of the user's real hand.
  3) As a feedback method, the bounding box of the selected object is shown.
C. We use the following filters as part of the GoGoIT:
  1) A MoveByLocator maps the user's hand position to the virtual hand or the cube -when the virtual hand is far away.
  2) A Select1ByTouching calculates which object collides with the hand representation.
  3) A ChangeObject shows/hides the bounding box of the selected object.
D. Some IT-specific filters are required:
  1) GoGoFilter: It receives as input the user's head and real hand positions and outputs the virtual position, according to the GoGo lengthening scheme.
  2) GoGoControl: It receives the user's hand position and the position calculated by the GoGoFilter and shows a cube, if they are different. It also activates/deactivates the MoveByLocator associated with the cube.
E. The detailed connection between filters is shown in Figure 3 as an object diagram in UML [4] [7]. Note that there are associations from the GoGoIT to the objects in the scene and the virtual input devices, but these objects are defined outside the IT [8].
F. Figure 4 provides a simplification of GoGoIT that shows only important details: input ports, output ports, and associations with objects in the scene [9].

---

[7] VIDevices and SceneObjects are packages that group related objects. Other rectangles represent objects. Directed lines are links between objects used to propagate messages. The diagram shows flow of information between filter's ports, flow of messages between filters and filters to objects, and filter grouping.

[8] In this way it is possible to change the objects related to an interaction technique, so different views and layouts are possible.

[9] handIport redirects the input to the three internal filters that require it, in Figure 3. gogoPosOPort comes from the GoGoFilter, and pickOPort comes from Select1ByTouching.

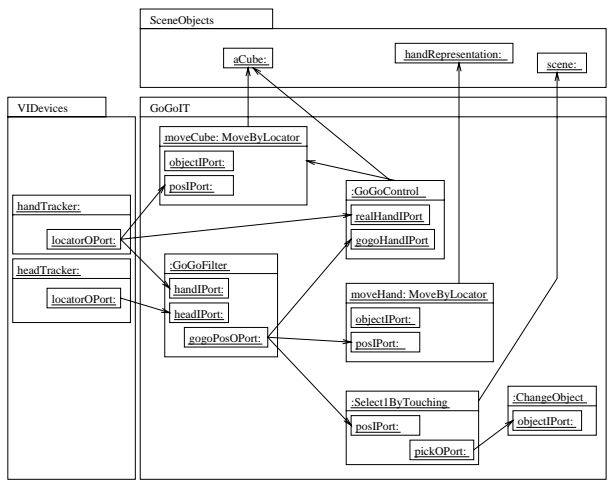

**Fig. 3** Representation of the GoGo Technique in ITLib.

Figure 5 shows a ray casting selection technique [17]. In this case, we need only the hand position and orientation as input. The hand representation and the selection ray are moved by the MoveByLocator filters. A Select1ByPointing filter selects an object from the scene and a ChangeObject filter provides the visual feedback.

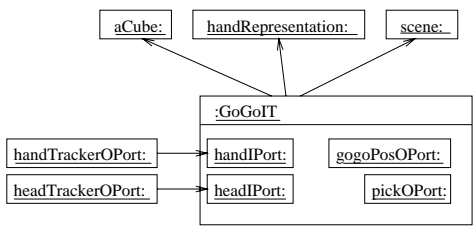

**Fig. 4** Encapsulated representation of the GoGo Technique.

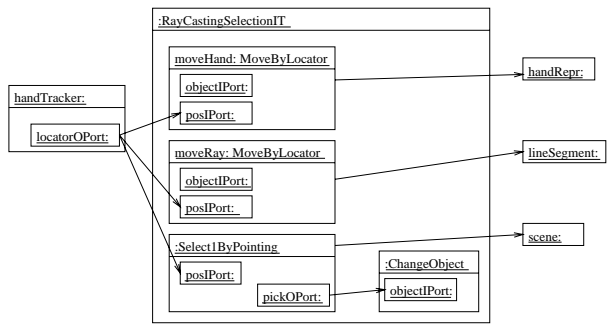

**Fig. 5** Representation of selection by ray-casting.

(2) Interchange of ITs

Interaction techniques and filters are combined to construct applications. For example, Figure 6 shows the design of an application to select and move an object that uses the GoGo Technique and two buttons to grab and release an object. GoGoIT sends the selected object and the filtered position to a MoveByLocator filter. MoveControl is an application-specific filter that switches between two modes (selection or movement) according to the input received from buttonGrab and buttonRelease.

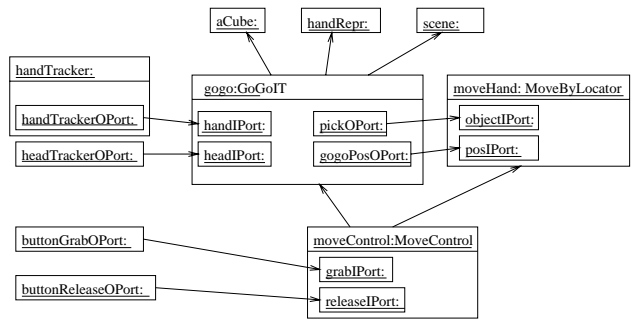

**Fig. 6** An application that moves objects, by using GoGoIT.

This application can change its selection behavior from GoGo to ray casting by sending the following messages (the final object structure is shown in Figure 7):

```
moveControl.off() // disables IT
gogo.disconnect()
handTracker.addLocatorListener( moveHand )
raycast.connect( handTracker, handRepr,
                 Scene, lineSegment, moveHand)
moveControl.setSelectionIT( raycast )
moveControl.on()
```

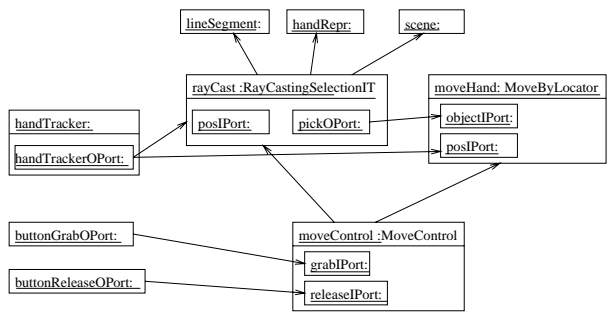

**Fig. 7** An application that moves objects, by using RayCastingIT.

### (3) Device Replacement

A device can be replaced with new ones or with a combination of others. For example, a 6DOF tracker can be replaced by 12 buttons (Figure 8), 2 for each dimension. Any IT can be plugged into this new device without any changes to the ITs code.

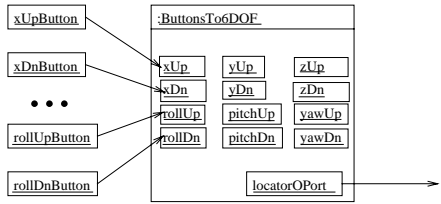

**Fig. 8** Device replacement.

### (4) A Walkthrough Application

An application that allows people to walk through the campus (Figure 10) has been redesigned using this model. The core functionality that allows a user to walk around designated areas thoughout the campus is shown in Figure 9. Input from the X window is received in an XInput object and transformed to information in its output ports, and a timer sends one event each execution of the dataflow. QuitByButton exists the application when a key is pressed, Motorcycle computes a new position according to the position of the mouse in the window, the window dimensions, and the buttons to start/stop the movement, and InsidePath decides if the new position is valid, according to a set of predefined paths. There is also a up/down movement computed by MoveUpDn that is combined with the result of InsidePath to define for each frame the final position of the viewpoint. This desktop application is being ported to a CAVE environment by replacing the XInput object, without affecting the rest of of the application. The framework allows us also to change the interaction technique for navigation, by replacing the filters Motorcycle and MoveUpDn for others, in order to evaluate several alternatives with the user.

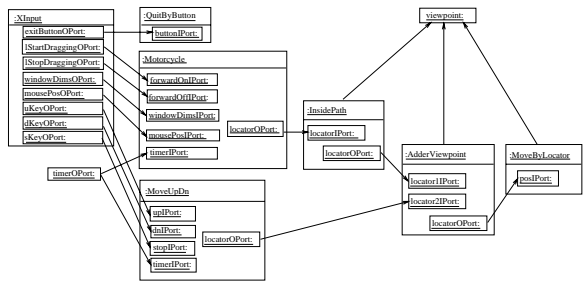

**Fig. 9** ITLib-based dataflow for the campus walkthrough.

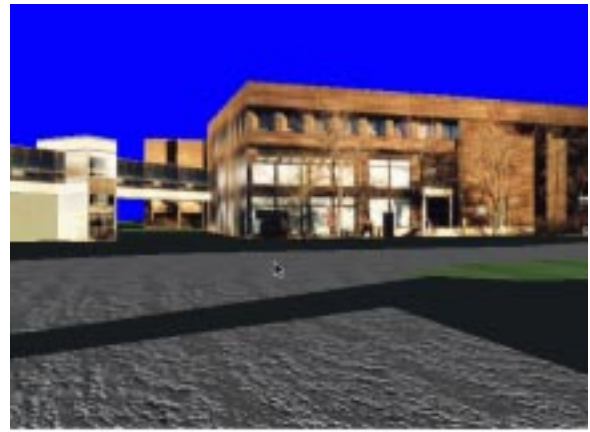

**Fig. 10** The campus walkthrough application.

## IMPLEMENTATION

We decided to implement a language-independent framework, due to the variety of environments we wanted to support. We chose XML [6] to describe ITs, because it is very easy to parse in order to generate code. Currently, The framework generates code for MRToolkit [25] (in C++), and porting to VRML and Java3D is in progress.

The main elements in the XML description are class, oport, and dataflowrel. A class is described in terms of patterns that define, at a higher level of abstraction, its attributes and methods. For example, the following class definition:

```
<class name="GoGoFilter" inherits="Filter">
  <prop name="center" type="Vector3D"
  access="r"></prop>
  <oport name="locator" type="OPort" >
  </oport>
</class>
```

defines the class GoGoFilter, a subclass of Filter, with a readable property called center and an output port called locator. These elements generate two attributes (_center and _oportLocator), and three methods (getCenter, addLocator Listener, and removeLocatorListener) in the C++ or Java implementations. A dataflowRel defines the interconnections between virtual input devices, objects in the scene and interaction techniques. For example:

```
<class name="cube">
  <object name="cube" type="Cube">
  <object name="viewpoint" type="Viewpoint">
```

```
<videv name="headTracker" type="MRLocator">
<it name="quit" type="QuitByNavigate">
<it name="moveViewpoint"
    type="Location2Viewpoint">
<dataflowRel origin="headTracker"
    port="locator" dest="moveViewpoint"
    port="iportLocator">
</class>
```

defines a virtual reality world that shows a cube, in which users can change the viewpoint with a tracker or quit the world. This generates a class and method calls in C++ or Java, or routes between objects and scripts in VRML.

The library's core is composed of 20 classes, and additional 44 classes define ITs, execution environments, input devices, and application-dependent filters for several examples we have developed. The relation between lines of code written using XML and the code generated in C++ is 1:1.47, and we expect to have better ratios once we add more applications and once we generate documentation.

## CONCLUSIONS

We have defined an architecture and a framework of interaction techniques for VR applications that can be plugged into existing VR toolkits. ITLib represents ITs as objects in an object-oriented library with abstraction, encapsulation, and extension mechanisms. Input and output devices can be added to the framework and used with pre-existing ITs, which are device independent. The execution model is a first class object that can be improved for better performance or real time constraints. The resulting environment allows us to setup an application by changing the ITs and devices we are using for equivalent ones, even at runtime. This work gives us a foundation to explore more complex 3D environments and 3D interaction techniques.

We plan to add more interaction techniques and execution models to ITLib, add support for multiuser environments, and create testbeds for virtual environments.